\documentclass[namedreferences]{solarphysics}
\usepackage[hyperref,optionalrh,solaromanenum]{spr-sola-addons} 
\usepackage{graphicx}                    
\usepackage{color}                       

\begin{document}

\begin{article}

\begin{opening}

\title{Photospheric magnetic field variations accompanying the 2011 June 7 eruptive event}

%
\author[addressref={iszf},corref,email={vfain@iszf.irk.ru}]{\inits{}\fnm{V.G.}\lnm{Fainshtein}}
\author[addressref={iszf},corref,email={egorov@iszf.irk.ru}]{\inits{}\fnm{Ya.I.}\lnm{Egorov}}
\author[addressref={iszf},corref]{\inits{}\fnm{G.V.}\lnm{Rudenko}}

%
\runningauthor{Fainshtein et al.}

\address[id=iszf]{Institute of Solar-Terrestrial Physics SB RAS, PO Box 291, Irkutsk, Russia}

\begin{abstract}
From vector measurements of the photospheric magnetic field with the SDO/HMI instrument, we studied the field variations within the 2011 June 7 eruptive event related to the filament eruption (FE), flare, and coronal mass ejection (CME). We analyzed the variations in modulus ($B$), in radial ($B_r$) and transversal ($B_t$) magnetic induction components, as well as in the field line inclination angle ($\alpha$) to the radial direction from the Sun center were. The conclusion is that the most probable cause for the FE onset being the trigger related to the flares and CME, as well as for Stage 1 of the filament dramatic acceleration, is «magnetic flux cancellation» in several sites inside and around the filament channel. We assumed that the flare and the stage of the filament fastest acceleration are related to a spot emergence near the south-eastern filament footpoint, in which the field polarity and value satisfy a possibility of this field magnetic reconnection with the field surrounding the filament. For the first time, we studied in detail the time variations in the field line inclination angles within different sites of the eruptive event. During the filament slow emergence, a decrease in the field line inclination angles was revealed to occur in its channel neighboring, whereas the field line inclination angles dramatically grew in the neutral line neighboring within the flare region after the flare onset. The flare ribbons at all the stages of their existence are shown to be located under the photosphere sites with the field modulus local maxima and with minima of the field line inclination angles. We show that, near the photospheric magnetic field polarity inversion line (PIL), the azimuth decreases after the flare onset, which means a decrease in the angle $\beta$ between PIL and the magnetic induction vector projection onto the sky plane. Far from PIL, on the contrary, the azimuth increases, and, correspondingly, the angle $\beta$ also increases.
\end{abstract}

%

\end{opening}

%
\section{Introduction}
     \label{S-Introduction} 

Solar eruptions (flares, filament eruptions (FEs), and coronal mass ejections (CMEs) are the brightest manifestations of solar activity accompanied by a release of great energy that is assumed to come from the solar magnetic field. Solar eruptions are considered to be accompanied by the magnetic field transformation within the eruption regions. This transformation, in the end, leads to the above events and determines the peculiarities of their evolution \citep{Forbes2000,Chen2011,Schmieder2013}.
To solar eruptions related are certain variations in the magnetic field at the level of its measurement (usually in the photosphere). New magnetic flux (NMF) emergence from under the photosphere is often considered drivers of solar eruptions are \citep{Feynman1995,Li2000,Sakajiri2004,Schrijver2009,Sun2012,Louis2015}. Eruptions may originate due to magnetic reconnection of the NMF fraction with a magnetic field certain polarity and with a sufficient field value with a site of the ambient field with the opposite polarity in the NMF location favorable for reconnection \citep{Chen2000,Archontis2008,Kusano2012,Leake2014}. In certain cases, reconnection occurs inside the new emerging magnetic arcade \citep{Archontis2010}.

Another peculiarity observed at the magnetic field measurement level that is closely related to eruptions is magnetic flux cancellation (MFC) \citep{Livi1989,Zhang2005,Welsch2006,Amari2010,Sterling2010,Green2011,Savcheva2012,Burtseva2013}. According to \citep{Livi1985}, the magnetic flux cancellation is «the mutual apparent loss of magnetic flux in closely spaced features of opposite polarity». \citealp{Welsch2006} offered a quantitative definition for MFC. Magnetic flux disappearance in closely spaced features with the opposite field polarity occurs, indeed, as long as the condition $\frac{\partial}{\partial t} \int dS |B_n| < 0$ is satisfied. Here, integrated is the field measurement level area involving the structures with the opposite polarity, in which the magnetic flux decreases (cancels) within a definite time $\Delta t$ by an equal value, $B_n$ is the magnetic induction component normal to the measurement field surface.

There were arguments that MFC occurs due to magnetic reconnection, and there were the cases addressed, when reconnection occurred below the field measurement level, on the latter, and above that level \citep{Zwaan1987,Yurchyshyn2001,Chae2003}. Based on both observations and MHD calculations, MFC was shown to b,e probably caused by ideal processes related either to the emergence of $U$-shaped field lines \citep{van2000,Magara2005}, or to the submersion of $\Omega$-loops \citep{Chae2004}.

Analyzing the relation between MFC and various solar phenomena (including eruptions) showed that the magnetic flux cancellation plays an important role in forming prominences \citep{Martin1998}, may be a trigger for a filament eruption and for CMEs \citep{zhang2001,Sterling2010}, and may accompany flares \citep{Burtseva2013}.

A physical reason for the MFC effect on eruptions is related to an increase (under certain conditions) in the free energy within the region containing the structures, where magnetic flux cancellation occurs \citep{Amari2003a,Welsch2006}.
In many cases, eruptions occur as follows: a filament (prominence) eruption - a flare - a CME \citep{Fainshtein2015}. Simultaneously, FE may be a trigger both for the flare and for CME \citep{Schmieder2013,Fainshtein2015}. To find the cause for the entire chain of eruptions, first of all, one should reveal the reason for a filament eruption. In literature, there are different FE mechanisms accounting for the field variation both inside and outside the filament \citep{Chen2011}. Some researchers relate FE to the emergence of a new magnetic flux (NMF) that often appears near the filament footpoint or near the filament channel, or, in certain cases, inside the filament channel \citep{Feynman1995,Sakajiri2004,Schrijver2009,Sun2012,Louis2015}. Nevertheless, in a number of papers, there is a conclusion that the NMF emergence is not the only condition for the eruption (CME) origin \citep{Zhang2008}. In certain cases, a filament eruption is related, as already noted, to the magnetic flux cancellation [Sterling et al, 2010]. Thus, until now, no one has established, what are necessary and sufficient transformations of the magnetic field at its measurement level that may precede the observed FEs, and also to accompany them, thereby reflecting the physical processes causing a filament eruption. Most likely, there are no such photospheric magnetic field variations (universal for all filaments) accompanying FEs, and different filament groups may be related to different variations in the magnetic field within the filament eruption region. This means that studying the photospheric field variations accompanying a filament eruption with different peculiarities continues to remain topical.

Magnetic reconnection is considered the basic mechanism for solar flares \citep{Priest2002}. Sharp and relatively large-amplitude jumps of the measured photospheric field obviously observed for sufficiently powerful flares after their onset indirectly reflect the 3-D magnetic field transformation occurring in this case (see \citealp{Kosovichev2001,Sudol2005,Petrie2010,Fainshtein2015}  and the references therein). But, what is the physical association between rapid variations in the magnetic field within the flare region in the corona and the field variations in the photosphere, has not been revealed until now. \citealp{Fletcher2008} assumed that the pulse of Alfven waves oscillated above produces the necessary forcing on the photosphere. In some papers, the authors endeavored to estimate the Lorentz force affecting the photosphere \citep{Hudson2008,Fisher2012}.

Besides the field variations within the NMF emergence region and sharp jumps of the field after the flare onset, within the region of the flare and of the other related eruptions, there occur, indeed, many other kinds of the eruption-related variations in the photospheric field, whose details have been studied poorly enough. For example, there are practically no data on the dynamics of the field line inclination angles within the filament eruption region and the related flare before and after its onset. In \citep{Hudson2008,Petrie2010}, there was an assumption that, within the flare region after its onset, an increase in the field line inclination angles should occur. \citealp{Petrie2010}, analyzing the amplitude of the magnetic field component jumps measured along the line-of-sight in the region of sufficiently powerful flares after their onset and originated at different distances from the central meridian, arrived at the conclusion that, on average, the field jump amplitude increases as the flares move from the central meridian. The authors of this paper assumed that such a result testifies to an increase in the field component transversal to the radial direction after the flare onset, and, as a consequence, an increase in the field line inclination angles to the radial direction. But this was not proven by direct measurements of angles.

In this paper, we present the results of studying the magnetic field variations accompanying the 2011 June 7 by using the vector measurements of the photospheric magnetic field at a high temporal and spatial resolutions. A special attention is paid to the variation in the field line inclination angles within the eruptive event region before and after its onset. We note that different aspects of this rather interesting event were repeatedly discussed earlier \citep{Li2012,Gilbert2013,Inglis2013,van2014,Dorovskyy2015}.

\section{Data Analysis Methods}
     \label{S-Data} 

We studied the magnetic field dynamics accompanying the 2011 June 7 eruptive event, including FE, flare, and CME. The characteristics of the photospheric magnetic field within the eruption region were found by using the field vector measurements with the Helioseismic Magnetic Imager (HMI; \citealp{Scherrer2012}) aboard the Solar Dynamic Observatory (SDO; \citealp{Pesnell2012}). The pixel size of the recording matrix is $\approx 0.5"$, the cadence being $\approx$ 12 minutes. Simultaneously, the field transversal component direction $\pi$-uncertainty problem originating in the field vector measurements was resolved through the technique suggested in \citep{Rudenko2014}. This method features a higher performance and accuracy of the problem solution, and is also applicable at larger distances from the solar disc center.

For the analysis, we used the field characteristics determined both in every pixel of the analyzed region image and averaged within the rectangles with the sides of 3 - 5 arc seconds and more, for $\approx$ 30 hours before the event onset and several hours after the latter. Analyzed were the variations in the modulus ($B$), in the radial ($B_r$) and in the transversal ($B_t$) magnetic induction components, as well as the variations in the field line inclination angle $\alpha$ to the radial direction from the Sun center. Practically, the $\alpha$ value was found from the relation: $cos(\alpha) = |B_r|/B$. The $B_r$ value was found by Relation (1) that included the measured values: $B$, angle $\delta$ between the field direction and the line-of-sight, and the azimuth, the angle $\psi$ measured on the sky plane counterclockwise from the CCD-matrix columns to the field transversal component that results when projecting the $B$ vector onto the sky plane. To determine the $B$, $\delta$, and $\psi$ parameters within different sites of the eruption domain, we used the Sun images with those parameters closest in time to the instants of obtaining Sun images in continuum.

Ratio (1) relates $B_r$ with the field components in the Cartesian coordinate system ($X$, $Y$, $Z$) to the center in the solar disk center, where the $OX$, $OY$ axes were placed on the sky plane, the $OY$ axis was directed to the North Pole (we neglected the angle between the ecliptic and the equatorial planes). The $OZ$ axis is perpendicular to the sky plane and is directed Earthward. We supposed that the line-of-sight is perpendicular to the sky plane in all the points within the solar disk. 

\begin{eqnarray}
  B_r=B_x X+B_y Y+ B_z Z=\nonumber \\
  B~sin (\delta)cos(\psi+90^\circ)+B~sin (\delta)sin(\psi+90^\circ)+B~cos(\delta) 
\end{eqnarray}

The angle δ between $B$ and $OZ$ varies within [$0^\circ$; $180^\circ$], the angle $\psi$ (azimuth) counted off in the sky plane from the $OY$ axis counterclockwise varies within [$0^\circ$; $360^\circ$].
The field transversal component was found from the relation: $B_t$ = $\sqrt{(B^2 - B_z^2)}$. Some analyzed sites of the active region involved sunspots. Their positions were determined from the Sun images obtained in the continuum with the SDO/HMI instrument. The Sun differential rotation on the solar images presented in the paper was compensated by the ``reprojection'' of the images to the same instant (2011 June 7 at 06:24 UT).
When building a time dependence of the eruptive filament front position, we used the 304 \AA -channel Sun images obtained with one of the Atmospheric Imaging Assembly telescopes (AIA; \citealp{aia}) aboard SDO. The one-dimensional time profiles for the eruptive filament position were obtained by using the 304\AA-channel Sun difference images along the direction at $\approx 10^\circ$ to the equator (southern latitude) in the western half of the solar disk and slit-averaged 2 pixels wide.

\section{Results}
     \label{S-Results} 

\subsection{What variations in the photospheric magnetic field might become a driver for the eruptive event?}

The 2011 June 7 event includes a large filament eruption, a flare, and CME formation, Fig. 1.

\begin{figure}[!ht] 
\centerline{\includegraphics[trim=0.0cm 0cm 0.5cm 0cm, width=1.00\textwidth]{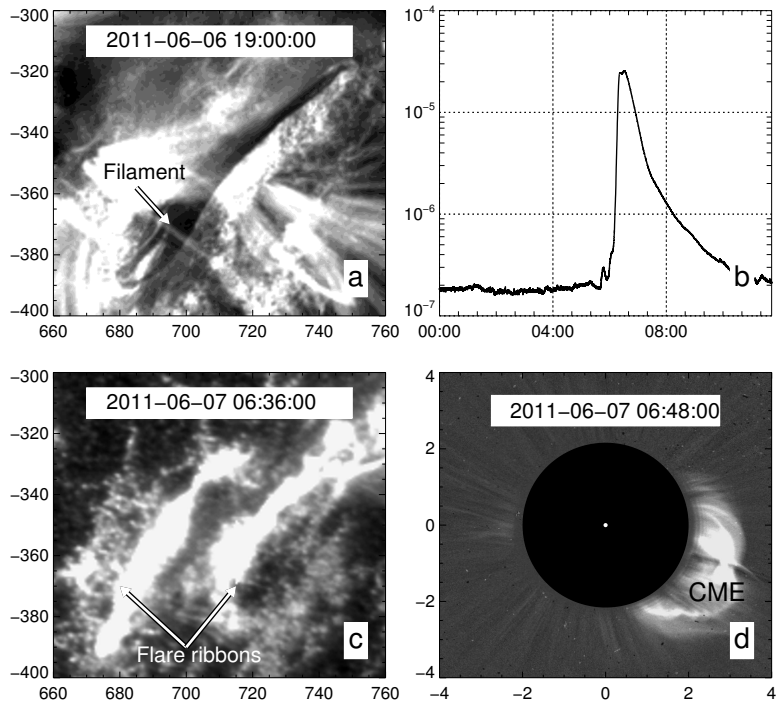}}
\caption{a) the filament during eruption from observations in the 171\AA~ channel; b) the soft X-ray flux within the (1-8)\AA~ wavelength range (GOES); c) a two-ribbon flare from observations in the 1700\AA~ channel; d) CME observed with the LASCO C2 coronagraph. }
\end{figure}

First of all, let us establish, in what sequence these eruptive processes occur. Fig. 2 shows the time variation in the eruptive filament position and velocity on the sky plane projection and marks the X-ray flare onset. One can see that the filament motion starts long before the flare. As illustrated in \citep{Fainshtein2015}, the filament eruption is the trigger for this CМE that forms after the flare onset.
On the eruptive filament velocity time profile, one can clearly see three periods of the velocity variation. For a long time, the eruptive filament velocity grows slowly, although with an intensifying acceleration, and at the instant indicated by the thick dotted line in Fig. 2 (a) reaches $\approx$ 0.35 km/s (Stage 1). From this moment (t$\approx$2011.06.07 (00:00)), a rapid growth in the velocity starts (Stage 2). But the greatest eruptive filament velocity jump occurred right after the flare onset (Stage 3). We endeavored to establish, whether the eruptive filament velocity variations at each of these stages relate to some magnetic field variations within the filament eruption region and in the adjacent sites of the active region. 

\begin{figure}[!ht] 
\centerline{\includegraphics[trim=0.0cm 0cm 0.5cm 7cm, width=1.00\textwidth]{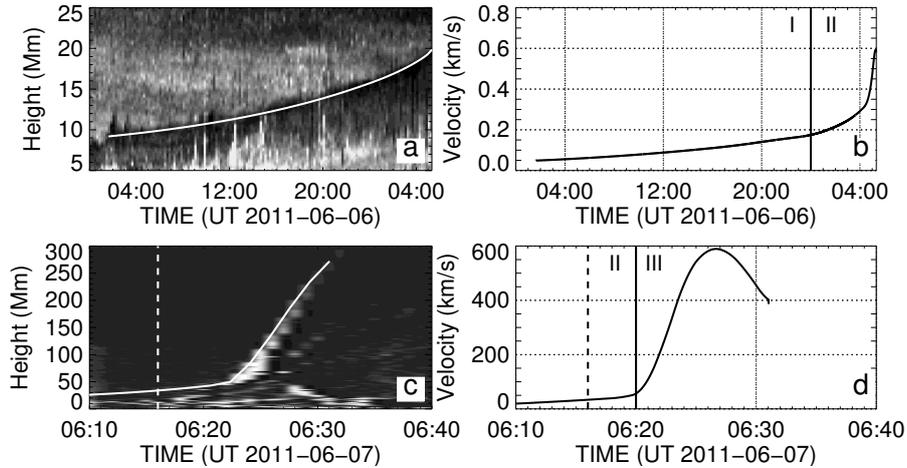}}
\caption{Time-dependant position and velocity of the eruptive filament boundary projection onto the sky plane: a) position, white solid line, b) velocity, black solid line, Stages 1 and 2 of the filament eruption; c) position, d) velocity, the Stage 2 end and the filament acceleration Stage 3. On the horizontal axes, the time in UT is counted off in hours from the 2011 June 6 beginning. Vertical solid thick lines mark the Stage 2 and 3 onsets, dotted thick line marks the flare onset. }
\end{figure}

We started analyzing the magnetic field variations accompanying the eruptions under study from $t_0 \approx$06.06.2011 (00:00). We managed to take the first measurement of the filament velocity at $t_1 \approx$ 06.06.2011 (02:00). The analysis of the magnetic field variations in NOAA 11226 inside and outside the filament channel showed that the strongest field variations occur in many sites, and these sites vary at different periods of the filament motion. We detected these sites through a visual continuum Sun image analysis. The addressed eruptions appeared to be accompanied by an intensive transformation of the active region sunspot pattern that manifests itself in the emergence and collapse of a great number of small sunspots and pores within different sites of the active region.

Figs. 3 (a, b, c, d) show the sites with the strongest variations in the sunspot pattern. Rectangles highlight them and, for reference convenience, the sites are numbered. Within each such a rectangle, there was an unsigned magnetic flux $\Phi = \int |Br|dS$, whose time dependence is shown in Figs. 3 (e, f, g, h). Let us address the $\Phi$ flux behavior peculiarities in Sites 1 - 8. Site 1 in the early 2011 June 6 is filled with a great number of pores. Eventually, the number of pores decreases, which is reflected in the $\Phi$ flux eventual decrease against minor variations. We note that Site 1 covers the regions with the opposite field polarity. And, although the magnetic flux $\Phi$ in the negative-polarity region is more than that in the positive-polarity region, the magnetic flux $\Phi$ variation rates in them are close. This allows us to draw a conclusion that, according to \citep{Welsch2006}, there is MFC in Site 1, and, consequently, these field variations may affect the filament motion, and, supposedly, be a trigger for the filament acceleration Stage 2.

\begin{figure}[!ht] 
\centerline{\includegraphics[trim=0.0cm 0cm 0.7cm 0cm, width=1.00\textwidth]{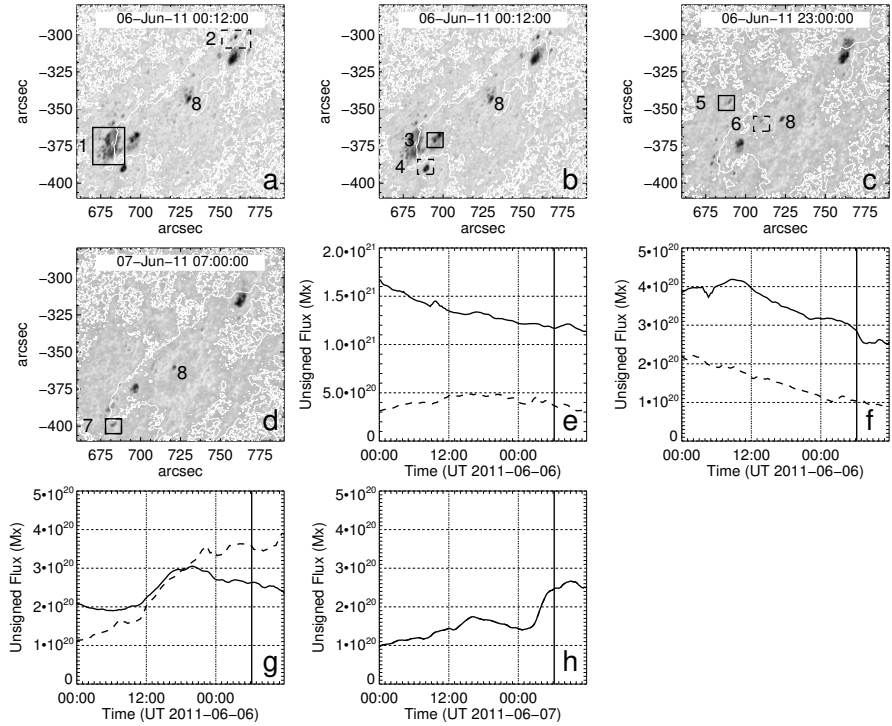}}
\caption{Sunspots around in the filament channel and in its close neighboring from the continuum observations in Sites 1-8 (a, b, c, d), and the time variations in the unsigned magnetic flux in Sites 1-7 (e, f, g, h). }
\end{figure}

The magnetic flux decrease with time (after the increase up to 14 UT 2011 June 6) was also observed in Site 2 located over the spot and covering the PIL fragments. There is also magnetic flux cancellation here in the opposite-polarity regions, starting with $t=14$ UT (2011 June 6).

Sites 3 and 4 involve the sunspots with the opposite field polarity. Starting with approximately 9:00 of 2011 June 6, the unsigned magnetic flux in both spots decreased at about the same rate. This means that the region combining these two spots meets the magnetic flux cancellation condition, i.е., may affect eruptive processes. In other words, the field variation in these spots also might become a driver for the filament acceleration Stage 2. We take notice of a sharp decrease in the magnetic flux in both spots after the flare onset. This is particularly evident in Spot 3. This means that eruptive processes may affect the sunspot magnetic properties of the active region, where eruptions emerge.

In unipolar Sites 5 and 6 with the opposite field polarity, there is a flux rise up to $t \approx$ 6/6/2011 (19 UT - 22 UT), and then, before the flare onset, or before the filament acceleration Stage 3 onset, the flux either decreases a little, or weakly varies. After the flare onset, a slight decrease in the flux is observed, but the value of this decrease is comparable with the amplitude of the preceding flux variations. Combining these sites into a large site allows us to consider that, in this combined site, MFC occurs starting with $\approx19$ UT - 22 UT. We note that the character of the magnetic flux behavior significantly varies with time in Sites 5 and 6 about 2-5 hours before the eruptive filament motion Stage 2 beginning at (t $\approx$ 2011 June 7 (00:00)), when the filament acceleration starts to increase dramatically (Fig. 2 (b)). This indirectly testifies to a possible role of the combined Sites 5 and 6 in the ``launch'' of the filament acceleration Stage 2.

Thus, we found that, in the eruptive filament channel neighboring (and, probably, inside it), there are some sites, in which, during different intervals before the flare onset, the different-polarity magnetic fluxes decreased, which, supposedly, reflects the magnetic flux cancellation in those sites \citep{Welsch2006}. Because such a process testifies to the existence, and, probably, to the accumulation of free energy in the above sites \citep{Welsch2006}, this energy may be spent to generate the filament eruption and the observed variations in the filament acceleration character. 

Shortly before the flare onset at $\approx$ 5:12 UT (2011 June 7) near the south-eastern footpoint of the filament, there appears a small spot, apparently, a pore, see Site 7 in Fig. 3 (d). Between $t=$5:24 and $t=$5:36, this spot dramatically displaced southward by about 4", and before the flare onset (at 6:12) expanded, and became darker. After the flare onset, within 5 hours, its size would increase and decrease, its brightness also varied. This spot peculiarity is that the value and the polarity of the field in it favored to the field magnetic reconnection in the spot with the ambient field outside the filament. Such reconnection might also contribute to the flare emergence and to the filament acceleration Stage 3.

Let us pay attention to another spot in Fig. 3 (Site 8). During all the observational time, the size and brightness of that spot varied throughout its evolution, particularly, after the flare onset, when the spot size and its brightness degree decreased distinctly. But the key feature for the Spot 8 evolution is the following. This spot displaced eventually south-eastward by $\approx$ 28" at, on average, $\approx 130~m/s$ starting with 00:00 (2011 June 6) before the flare onset. After the flare onset, it moved by $\approx 2"$ for $\approx$ 5 hours, but at a smaller velocity. This displacement correlates with the motion (in the same direction) of the filament south-eastern footpoint and the adjacent magnetic field PIL bending (see Fig. 5). Such a filament footpoint displacement probably reflects an increase in the filament shear, which itself may produce the conditions for the filament disbalance and for a sharp increase in its velocity at one of the instants that we addressed above (Fig. 2). Unfortunately, we failed to analyze a similar motion of the north-western footpoint in the opposite direction due to the effect of the ambient external structures that did not allow us to determine that footpoint position within the necessary precision. From the aforesaid, it follows that one may regard the Spot 8 motion before the flare onset or before the filament velocity increase start (close to this time) as an indicator for the filament shear increase.

\begin{figure}[!h!t] 
\centerline{\includegraphics[trim=0.0cm 0cm 0.7cm 4cm, width=1.00\textwidth]{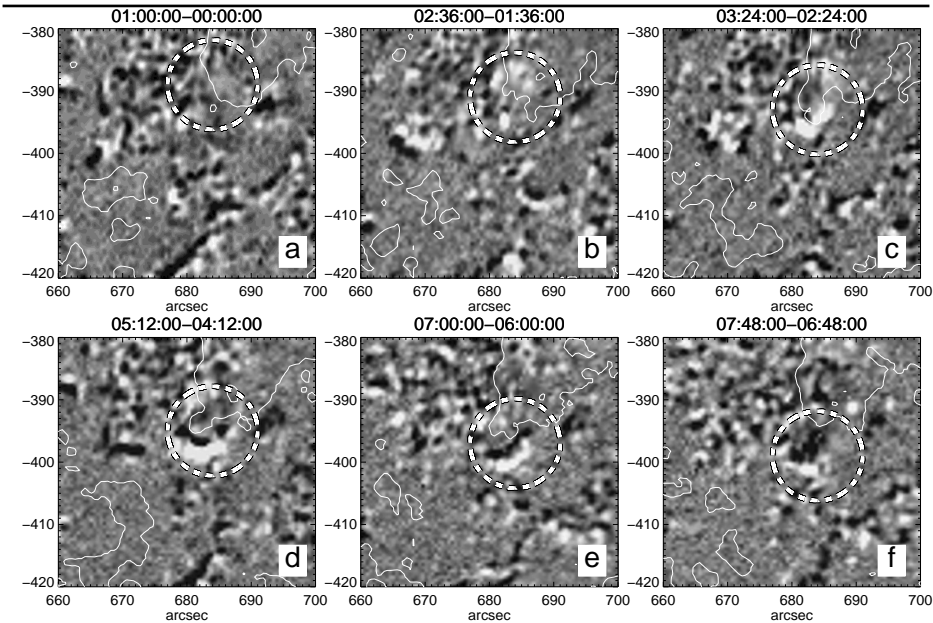}}
\caption{Example of rapid variations in the magnetic field near the south-eastern filament footpoint on difference images of the magnetic induction modulus (the difference between the closest images was 1 h) before the flare onset. The domains of the strong variations in the magnetic induction are encircled.}
\end{figure}

Thus, here, we only assume, what variations in the photospheric magnetic field in NOAA 11226 may be related to the filament eruption and the flare (we consider the CME formation as an FE consequence). Those are: (1) magnetic flux cancellations in several sites, (2) sunspot origin near the eruptive filament footpoint. The value and the polarity of this spot field favor the magnetic reconnection of the spot field with the field surrounding eruptive filament; (3) filament shear increase along the photospheric field neutral line before the flare onset or before the FE Stage 3 start close in time.

In NOAA 11226, we failed to find the emergence of long-living, relatively large-size, new bipolar magnetic fluxes similar to those observed before the filament eruption in (Feynman and Martin, 1995). Nevertheless, the NMF emergence as a pore group was observed in Sites 2, 5-6. But, in all the cases, starting with a certain instant, the magnetic flux in these sites decreased, at least, before the flare onset. Note that, in several sites within the filament channel field and in its neighboring, regions of short-living (12 min - several hours) new magnetic fluxes were observed before the flare onset. Figure 4 shows the origin of NMF near the PIL south-eastern bending next to the corresponding footpoint of the eruptive filament. We chose $t_0 = 01:00$ UT (2011 June 7) as the initial moment for timing. At $t_2$ = 02:48 UT (2011 June 7), preferentially in the field positive-polarity domain, observed was the field increase that manifested itself in emergence of white magnetic hills. During $t_3$ = 03:00 UT (2011 June 7), the field increased within one hill in the positive-polarity domain (right of PIL) and within another hill in the negative-polarity domain (left of PIL). At $t_4$ = 03:24 UT (2011 June 7), the hill of a higher field induction within the positive-polarity domain practically disappeared, and, further, during $t_5$ = 03:36 UT (2011 June 7) and $t_6$ = 04:24 UT (2011 June 7) it was not observed. But, simultaneously, during these instants, the field in the arrow-marked magnetic hill in the negative polarity domain continued to increase. This hill size also increased. Note that this magnetic hill displaced eventually with the PIL bending south-eastward. Later, in place of this magnetic hill, there appeared Spot 7, whose possible role in generating the flare we have already discussed.

\subsection{On some peculiarities of photospheric magnetic field variations within the eruptive event region}

Fig. 5 shows the distribution of the magnetic induction modulus ($B$) in NOAA 11226 and in some sites of the closest neighboring for several instants. PIL is also plotted on these distributions. It appeared that, small-scale and relatively rapid (timescale varying from 12 minutes to several hours) field variations occurred in the addressed region within $\approx$30 hours before the flare onset. The main peculiarity of the time variation ($B$) before the flare onset is in splitting the extended A domain with the magnetic field higher values into a set of relatively minor fragments with a magnetic induction higher value, and with a noticeable transformation of $B$ and C domains. One may consider this transformation as an assemblage of origins and disappearances of the magnetic hills with the magnetic field higher values. In the $B$ domain, one observes the eventual motion of some field sites south-eastward. One of such sites includes Spot 8, the other involves the domain with Spots 6. Before the flare onset, the sunspot and the spot domain start to split, and, several hours after the flare, they finally diverge. The observed field variations within the A and $B$ domains indirectly agree with a previous conclusion that MFC occurred in these domains. With time, the C hill domain distorts most complicatedly. Eventually, it disappears, but an extensive domain of higher $(B)$ values forms near due to occurrences of new hills and travel of the already existing ones.

\begin{figure}[!ht] 
\centerline{\includegraphics[width=1.00\textwidth]{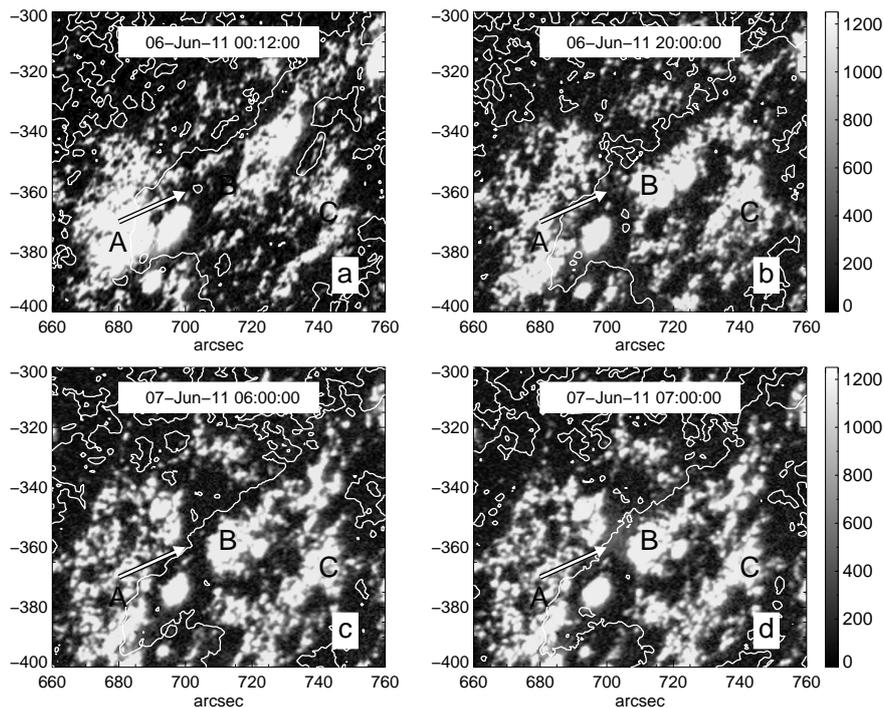}}
\caption{Distribution of the magnetic induction modulus in the 2011 June 7 eruption region at different instants. The arrow shows the domain of a noticeable increase in the magnetic field after the flare onset.
}
\end{figure}

Note the domain around PIL resembling the letter X that has the center with the coordinates x $\approx$ 705", y $\approx$-365". Right after the flare onset, the brightness in this domain starts to vary: this domain color varies from black to gray, which evidences an increase in the magnetic induction modulus $(B)$ here. Later, we will discuss how great this increase is. Here, nipping on ahead, we note that the observed magnetic induction increase is mainly caused by an increase in the magnetic field transversal component $B_t$ within this domain. Fig. 6, in which the ($B_t$) increase domain becomes lighter, illustrates this increase in the field transversal component. As to the ($B_t$) variation within the A, B, and C domains highlighted on the $(B)$ distribution, the field transversal component in the A domain decreased, on average, before and after the flare onset, and, only around a small PIL segment, the transversal component increased. On the contrary, in the $B$ domain, this component eventually increased, on average, at the segment corresponding to Domain 6 on the continuum images, whereas, in the C domain, in different sites, the ($B_t$) value varied chaotically.

\begin{figure}[!ht] 
\centerline{\includegraphics[width=1.00\textwidth]{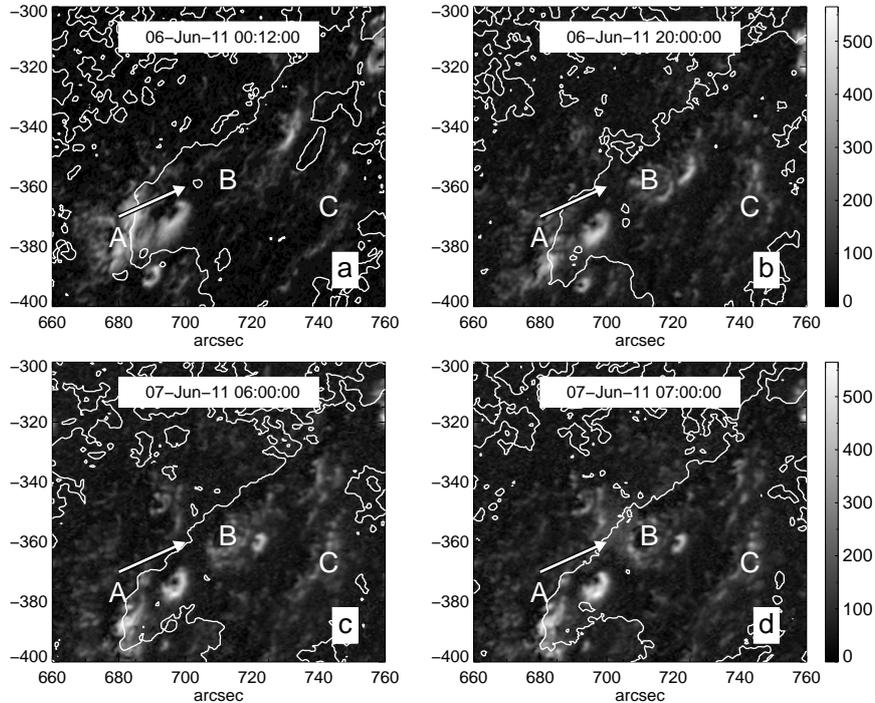}}
\caption{Magnetic induction transversal component distribution in the 2011 June 7 eruption region at different instants. The arrow shows the domain of a noticeable increase in the field transversal component after the flare onset.}
\end{figure}

Fig. 7 illustrates the variation in the field line inclination angle to the radial direction from the Sun center ($\alpha$) during the period under analysis. Before the flare onset ($t_5$ = 06:16 UT (2011 June 7)), the domain occupied with relatively large inclination angles generally decreased with time on panels (a-c). This manifests itself in the white domain decrease. The bulk of the panel area becomes tighter. Such a decrease in the angles is clearly observed in the neighboring of the PIL extended segment with the coordinates of its ends being [x=680", y =-390"]; [x=690", y =-355"], Fig. 7 (a-c). At the outset of our observation period, the angle ($\alpha$) reached maximal values within a wide strip around PIL. Eventually, this PIL segment distorted, and the domain of angle $\alpha$ great values around this PIL segment dramatically decreased in the most part of the above PIL segment. In other sites of the active region, the greatest $\alpha$ values appear, as one may expect, around numerous PILs, and the variation in the angle $\alpha$ spatial distribution eventually traced the variation in the distribution of PIL fragments within the active region.
Also, the field line inclination angle ($\alpha$) to the radial direction from the Sun center appeared to increase within the $B$ and $B_t$ increase domains after the flare onset, Fig. 7 (d) (the arrow shows the increase domain).

\begin{figure}[!ht] 
\centerline{\includegraphics[width=1.00\textwidth]{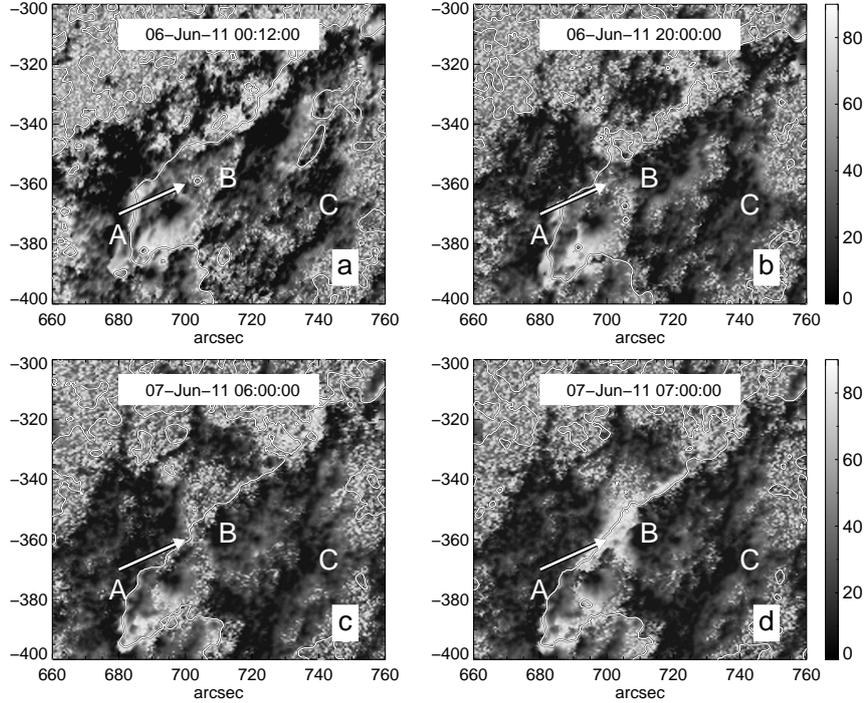}}
\caption{Distributions of the field line inclination angles ($\alpha$) to the radial direction from the Sun center for several instants. The arrow shows the domain of a noticeable $\alpha$ increase after the flare onset.}
\end{figure}

We noted above that in a number of papers \citep{Hudson2008,Petrie2010}, there was an assumption of an increase in the field line inclination angles within the flare region after its onset. Fig. 7 corroborates this. Note only that, in the addressed event, the growth in the angles was observed only over a part of the flare region (cf. the angle $\alpha$ increase domain with the entire flare region in Fig. 1(c)).

Next figures present the peculiarities of time variations in B, $B_t$, $B_r$, and $\alpha$ within the NOAA 11226 individual sites. Fig. 8 illustrates the $B$ time variations in several squares located along PIL in the domain of $B$ values increased during the flare, and displacing relative to PIL at different distances (Fig. 8 (a) shows the positions of the squares superimposed on the magnetic induction modulus distribution at $t=06:24:00$ UT). One can see that while the squares are within the domain of $B$ values increased during the flare, after the flare onset, observed was a relatively rapid growth in $B$ by $\approx$ (130 - 160) $G$, persisting against slight variations for, at least, 6 hours. Before the flare onset, in the squares located along PIL, the magnetic induction modulus differed, but, eventually, at least, from 16:00 UT (2011 June 6), $B(t)$ decreased in all the squares. Several hours before the flare onset, the field became approximately equal (142 $G$) in all the squares, and slightly varied within $\approx$5 hours. Then, the $B$ jump decreased after the flare onset, and, finally, disappeared, as it moved to the boundary of the domain with the magnetic field increased after the flare onset and when it exited this domain (Figs. 8 (b, c. e, f)). The character of the $B(t)$ variation before and after the flare onset substantially differed in different squares remote from PIL, and one can hardly reveal common regularities in the $B(t)$ behavior in these squares.

\begin{figure}[h!t] 
\centerline{\includegraphics[trim=0.0cm 0cm 0.7cm 8cm, width=1.00\textwidth]{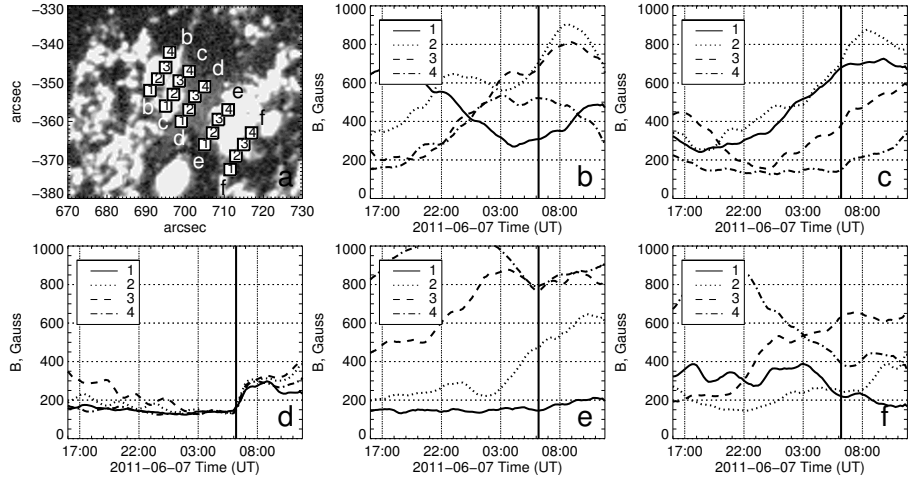}}
\caption{Panel (a) shows the positions of sites in NOAA 11226 (the squares are numbered, whereas the rows of squares are lettered) in the domain of the magnetic induction elevated values B, and in the closest neighboring of this the domain) plotted on the $B$ directional distribution at t$=06:24:00$ UT (2011 June 7). Panels b, c, d, e, f present the time variation in the magnetic induction modulus $B$ within different squares. Panel g exhibits the position of several sites arbitrarily selected outside the domain, in which the $B$ value grew after the flare onset. On the $B$ distribution, arrows show the domain, where the magnetic induction grows after the flare onset. Panel h provides the $B(t)$ time variations in these squares.}
\end{figure}

Figs. 9 shows the dependences similar to those in Figs. 8, but for the transversal component of the magnetic induction $B_t$. One can see that the character of the $B_t(t)$ variation is similar to the $B(t)$ behavior within the squares located along PIL in the domain of the magnetic field growth after the flare onset. This means that the growth in the magnetic induction after the flare onset is mainly caused by the increase in the field transversal component. Before the flare onset, $B_t(t)$, as well as $B(t)$, was approximately equal in the (d) squares for several hours, and slightly varied with time. But before this period, the character of $B_t(t)$ variation was different in different squares, and differed from $B(t)$ in the same squares. In the b, c, e, f squares, the $B_t(t)$ behavior distinctly differed, but one can see the following regularity. When there is even a slight $B_t$ growth in a square after the flare onset, a slight variation in the field transversal component with a period from several hours to $\geq14$ hours preceded the flare onset.

\begin{figure}[h!t] 
\centerline{\includegraphics[trim=0.0cm 0cm 0.7cm 8cm, width=1.00\textwidth]{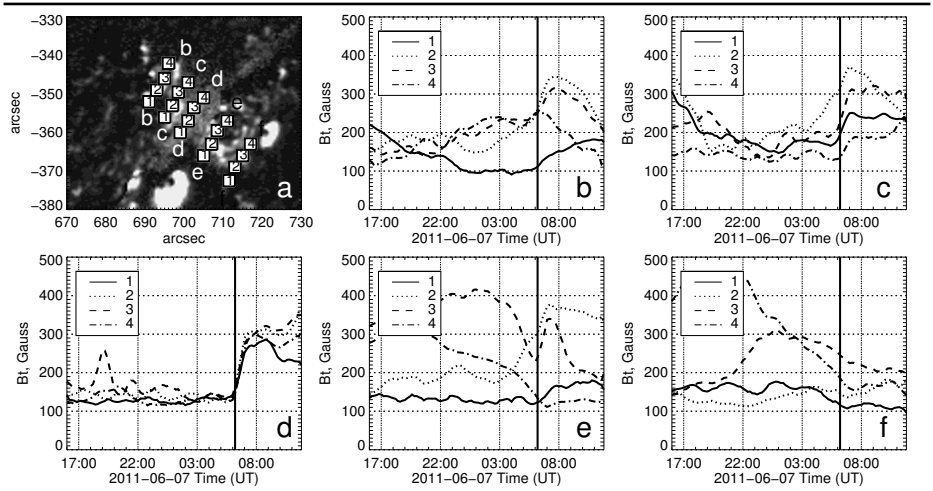}}
\caption{The same as in Fig. 8, but for the magnetic induction transversal component.}
\end{figure}

Fig. 10 illustrates the time variations in the magnetic induction radial component $B_r$ in the same sites of the eruption region. Note some peculiarities of $B_r$ time behavior. In the on-PIL squares, before the flare onset, $B_r$ weakly varied with time, and, for about 5 hours before the flare onset and 2 hours after the flare onset, $B_r$ $\approx$ 0. In other squares, the $|B_r|$  variation had a different character before and after the flare onset. In some cases, the flare was preceded by the $|B_r|$  growth, in others, it was preceded by the $|B_r|$  decrease, and, still in others, $|B_r|$  weakly varied with time. A similar difference in the $B_r(t)$ behavior in different squares was observed after the flare onset.

\begin{figure}[h!t] 
\centerline{\includegraphics[trim=0.0cm 0cm 0.7cm 8cm, width=1.00\textwidth]{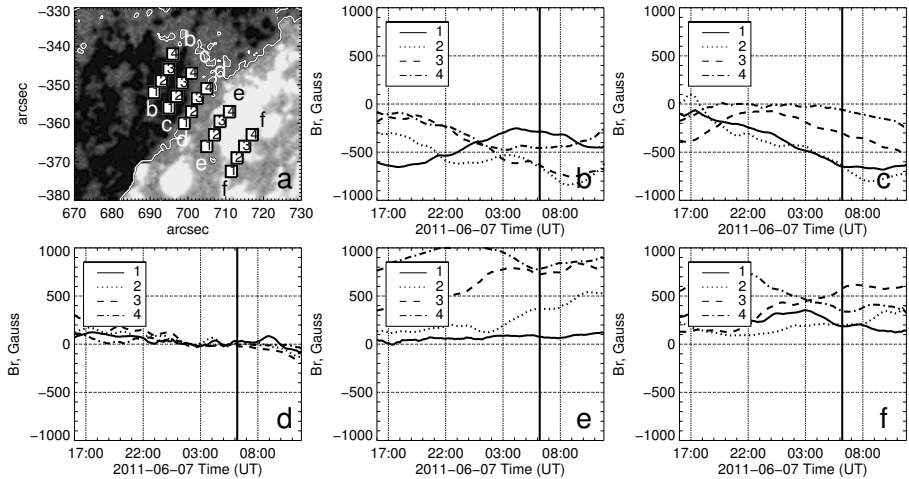}}
\caption{The same as in Figs. 8, but for the magnetic induction radial component.}
\end{figure}

Figs. 11 (b, c, d, e, f) show the field line $\alpha$ angle inclination variations averaged within the same squares (panel a), as those in Fig. 8. In this figure, the location of squares is shown against the distribution of $\alpha$ angles in the eruption region at the instant about half an hour after the flare onset. We start our analysis with the variation in angles after the flare onset. One can see that in the on- PIL squares (this domain is colored white), the $\alpha$ angle, after the flare onset, grew relatively dramatically with time, increasing by $\approx30^\circ - 33^\circ$, and reaching maximal values of $\approx83^\circ - 89^\circ$ after about 40 minutes. Upon reaching the maximal value, at least within 6 hours, $\alpha$ either weakly varied or decreased against the variations, but continued to remain beyond the values that were at the flare onset. Note that the angle $\alpha$ growth started not with the flare onset, but 5-25 minutes before that time. At the flare onset, the $\alpha$ growth rate changed in some squares.

\begin{figure}[h!t] 
\centerline{\includegraphics[trim=0.0cm 0cm 0.7cm 8cm, width=1.00\textwidth]{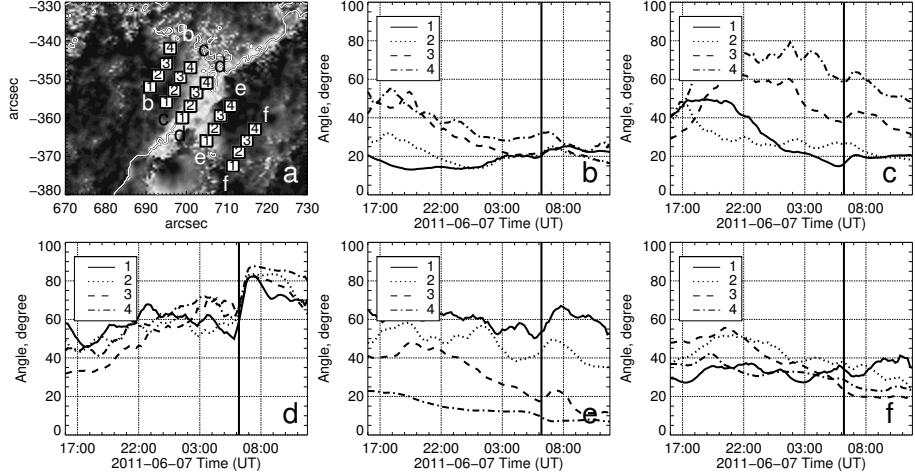}}
\caption{a - h: the same as in Figs. 8, but for the field lines inclination angle $\alpha$.}
\end{figure}

We may explain the angle values in the on-PIL squares before the flare onset, and its variation after the flare onset in the following way. Although exactly on PIL, where $B_r=0$, the angle $\alpha = 90^/circ$, in the on-PIL square, there occurs the field line inclination angle averaging, and, due to a rapid decrease in $\alpha$ within the square as the latter moves from PIL, the average $\alpha$ angle within the square appears to be significantly less than $90^\circ$. The $\alpha$ angle growth in the squares after the flare onset means that the $\alpha$ drop rate decreases during this period when moving from PIL. The comparison among the $\alpha$ angle distributions close to PIL in the domains, where $\alpha$ increased or practically did not vary after the flare onset (Figs. 8 (a, c, e, g, i)) qualitatively corroborates this conclusion.

As the squares, where the angles $\alpha$ is averaged, move away from PIL (Fig. 11(a), rows of squares b, c, e, f), the $\alpha$ jump decreased during the flares, and, in the end, it either disappeared or became very small. 

The character of the angle $\alpha$ variation before the flare onset depends on the distance of the squares, where the angles $\alpha$ values are averaged, from PIL. In case, when these squares are on PIL, an increase in the angles $\alpha$ against the variations before the flare onset over a period from several to 30 hours is characteristic. At large distances from PIL, angle $\alpha$ decreased against the variations in the most squares before the flare onset, with the periods varying depending on the square.

In the squares located outside the domain, where angles $\alpha$ grew after the flare (Fig. 12(left panel in the upper row, Squares 1-4)), including the PIL squares (Square 3), the angles decreased against the variations, both before the flare onset and after it (Fig. 12(right panel in the upper row). At the same time, one can see that, in details, the $\alpha$ angle time variations significantly differ depending on the square position. We note that starting with t $\approx$ (2 - 3) UT 2011 June 6, the $\alpha$(t) decrease rate dramatically increases in all the squares, and this instant appeared within the initial stage of the eruptive filament velocity first sharp change (Fig. 2 (b)).

Besides analyzing the $\alpha$(t) behavior in small-size localized sites of the photosphere, we analyzed the time variation in the $\alpha$ angle (Fig. 12(right panel in the lower row) averaged within relatively large sites of the photosphere (Sites 5-8), Fig. 12(left panel in the lower row). One can see that, in Rectangles 5-7, the $\alpha$ angle started decreasing from t$\approx$00:00 UT (2011 June 7) (the start of the filament dramatic acceleration Stage 1), and in Rectangle 8 this process started at t$\approx$03:00 UT (2011 June 7). After the flare onset, the angle $\alpha$ continued decreasing for $\approx$1 - 4 hours, and, then, it either weakly varied or grew. Note that both in the small squares, and in the large rectangles located outside the domain of the angle $\alpha$ growth after the flare onset, the character of the $\alpha$(t) variation significantly changed at the flare onset.

\begin{figure}[h!t] 
\centerline{\includegraphics[trim=0.0cm 0cm 0cm 1cm, width=1.00\textwidth]{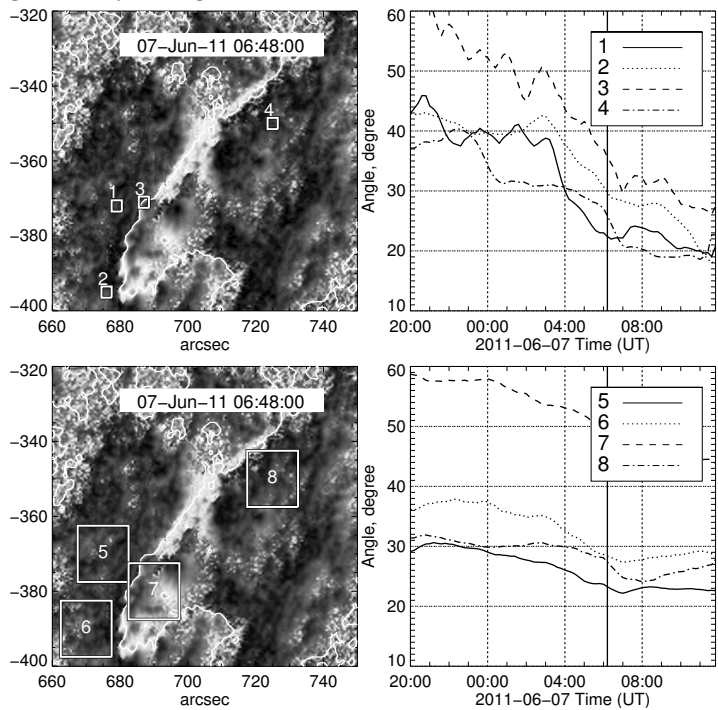}}
\caption{Upper row: left panel shows the spatial distribution of the field line inclination angle $\alpha$ after the flare onset with plotted Squares 1-4 outside the domain, where the angles $\alpha$ increased after the flare onset. Right panel presents the time dependences for the angles $\alpha$ values averaged within Squares 1-4. Lower row: left panel provides large-size Rectangles 5-8 superimposed on the $\alpha$ angle spatial distribution after the flare onset. Right panel exhibits the time variations in the $\alpha$(t) values averaged within each large rectangle.}
\end{figure}

The angle $\alpha$ appeared to increase before the flare onset in a small region near the eruptive filament south-eastern footpoint (we do not yield the $\alpha$(t) dependence for this case). One may elucidate such a behavior of the field line inclination angle by a departure of the field lines from the vertical direction under the effect of the filament footpoint displacing, as already noted, south-eastward during the filament eruption. 
\begin{figure}[h!t] 
\centerline{\includegraphics[trim=0.0cm 0cm 0.7cm 4.5cm, width=1.00\textwidth]{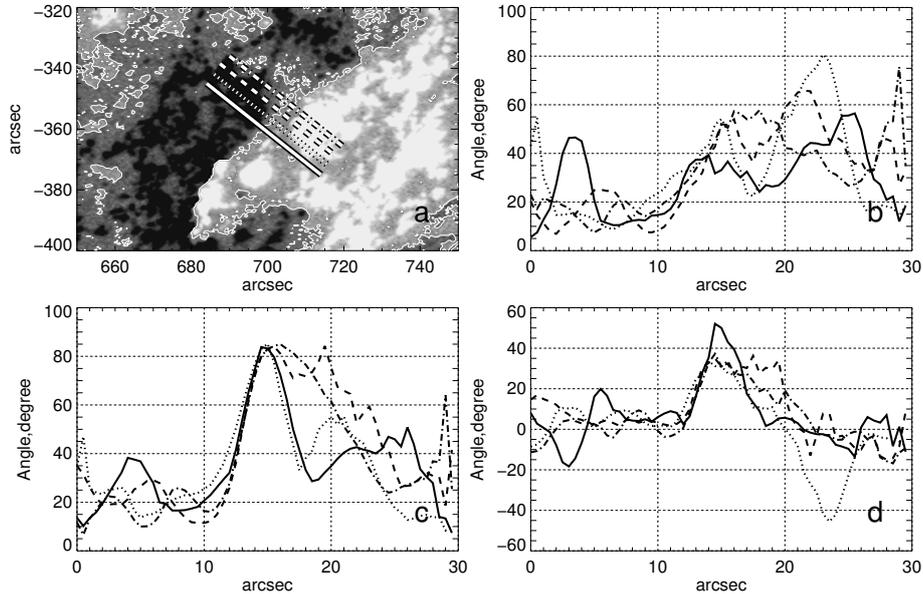}}
\caption{Panel (a) presents the lines crossing PIL. It is these lines, along which we obtained the distributions of the field line inclination angles $\alpha$ before the flare onset at $t_1$  = 05:00 (Panel (b)) and after the flare onset at $t_2$  = 07:00 (Panel (c)). Panel (d) shows the angle $\alpha$ distribution difference at $t_2$  and $t_1$.}
\end{figure}

Figs. 13 (a - d) allow us to estimate the spatial scale of the angle $\alpha$ variation relative to PIL. Fig. 13 (a) shows the lines crossing PIL in several points and the domain of angles $\alpha$ elevated during the flare in the PIL neighboring. Figs. 13 (b-d) presents the $\alpha$ distributions along these lines at $t_1$  = 05:00 (before the flare onset) and $t_2$  = 07:00 (after the flare onset), as well as the angle $\alpha$ difference distribution along these lines. One can see that the angle $\alpha$ difference distributions along these straight lines differ for each straight line, but on all the straight lines, within approximately 25", angles $\alpha$ at $t_2$  are more than at $t_1$.

As stated above, the flare accompanying this eruptive event is a two-ribbon one (Fig. 1 (C)). Flare ribbons are regions of increased emission intensity. They are observed in chromospheric lines, in the transition zone lines, and in some ultraviolet lines of the corona as regions of increased emission intensity \citep{Emslie2003}. For sufficiently powerful flares, the number of ribbons is usually 2, but there are many flares, in which a larger number of ribbons is observed. In two-ribbon flares, the ribbon central regions are usually parallel, but, generally, flare ribbons may have diverse shapes. In eruptive flares, the ribbon shape resembles the letter J. Flare ribbons are considered to be feet of flare loops.

It is interesting to establish, what peculiarities of the photospheric magnetic field under flare ribbons are. Flare ribbons are known to eventually diverge, moving away from the neutral line \citep{Hudson2011,Chen2011c}. What occurs to the photospheric magnetic field in the ribbon region in this case? Figs. 14 and 15 answer this question. The first of these figures shows the angle $\alpha$ distributions along the lines crossing the flare ribbons. We present these lines for different instants with different locations of the flare ribbons on the left panels in the figure. The crosses on these lines mark the points, where angles $\alpha$ reach local minima. For all the locations of the flare ribbons, these crosses appeared within the ribbons. The vertical thick lines transit through the crosses on the straight lines. From this figure, it follows that the flare ribbons are over the photosphere regions with the angle $\alpha$ local minima. In other words, the field lines crossing the photosphere under the flare ribbons appear more vertical, than those not crossing the flare ribbons.

\begin{figure}[h!t] 
\centerline{\includegraphics[trim=0.0cm 0cm 0.7cm 1cm, width=1.00\textwidth]{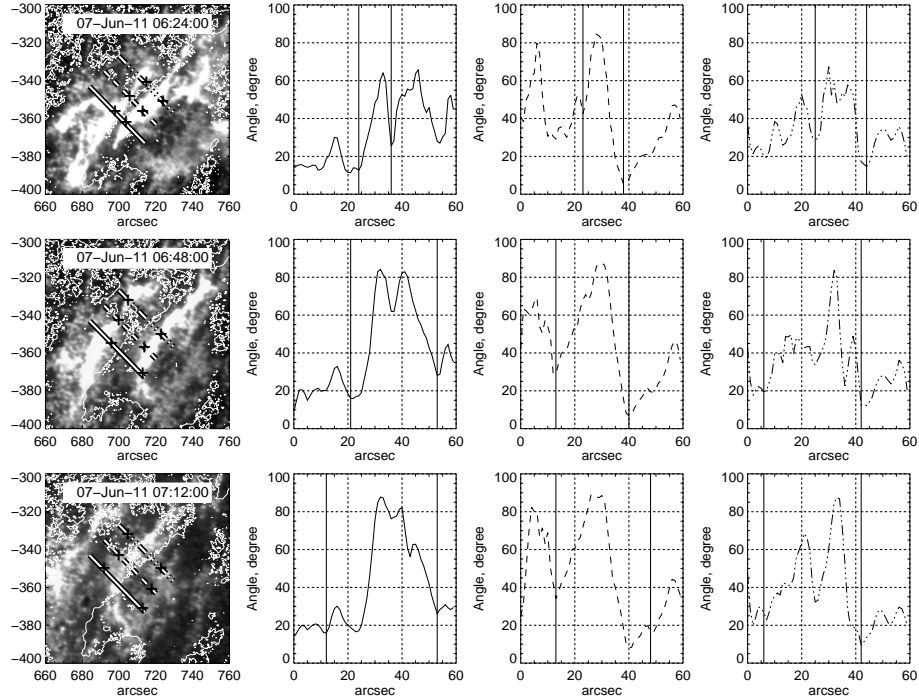}}
\caption{Left panels show images of the Sun fragments in the 1700 \AA line at different instants. On these panels, the flare ribbons are outlined (white fringes), and plotted are the lines along which the field line inclination angles $\alpha$ were scanned. Right of the Sun images, shown are the angle $\alpha$ distributions along these lines. The distance along the lines is counted off from their top ends. The kinds of lines (solid, dotted, etc.), along which the angles $\alpha$ were scanned, and of those on the plots coincide. }
\end{figure}

Fig. 15 exhibits the magnetic induction modulus $(B)$ distributions along the straight lines shown on the left panels in Fig. 15. These lines coincide with the straight lines, along which there were the angle $\alpha$ distributions. One can see that the flare ribbons are over the photosphere sites in the neighboring of the $B$ local maxima.

\begin{figure}[h!t] 
\centerline{\includegraphics[trim=0.0cm 0cm 0.7cm 1cm, width=1.00\textwidth]{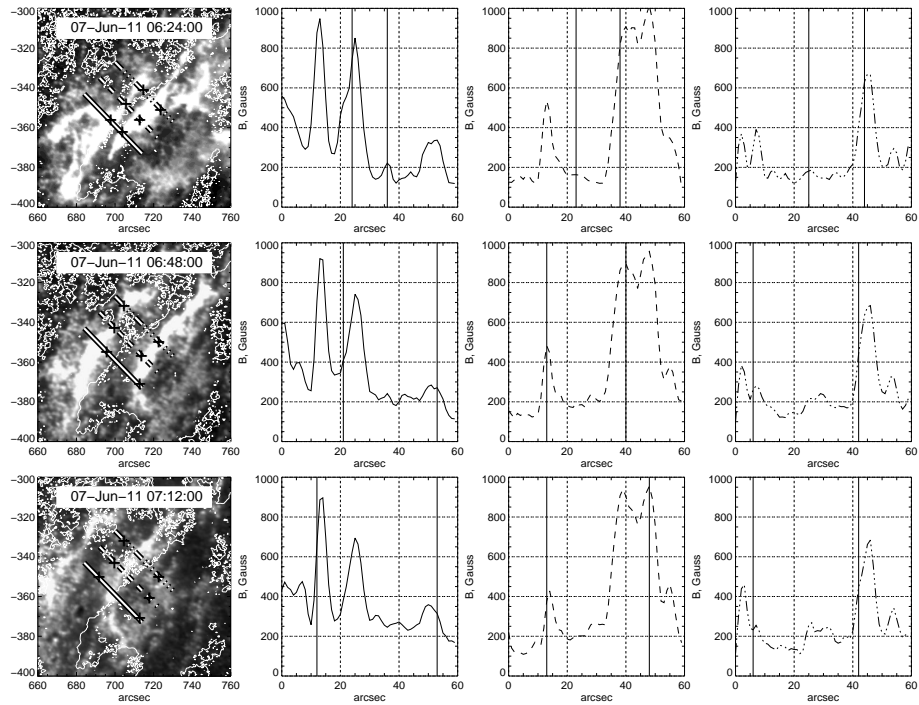}}
\caption{The same as in Fig. 14, but for the magnetic induction modulus $B$.}
\end{figure}

The issue of the magnetic induction vector azimuth $\psi$ variation before and after the flare onset is also interesting. The answer to this issue allows us to reveal, in particular, how the orientation of the magnetic field lines encompassing PIL (and the filament channel, as well) changes relative to PIL. From Fig. 16, it follows that the azimuth time variation character depends on the position of the square, in which this change is analyzed. Practically, we failed to reveal any regularities for the $\psi(t)$ variation that are inherent in all the squares, where the field azimuth is averaged, before the flare onset. In the on-PIL squares, the characteristic trend (before the flare onset) is the azimuth decrease with time against the large-amplitude variations. After the flare onset, in the on-PIL squares, one observes a dramatic decrease in the angle $\psi$. This means that the angle between the magnetic induction vector $B$ projection onto the sky plane and PIL decreases. In the squares most remote from PIL, after the flare onset, one observes a slight (up to $\approx 10^\circ$) azimuth increase, which means an increase in the angle between PIL and the vector $B$ projection onto the sky plane.

\begin{figure}[h!t] 
\centerline{\includegraphics[trim=0.0cm 0cm 0.7cm 8cm, width=1.00\textwidth]{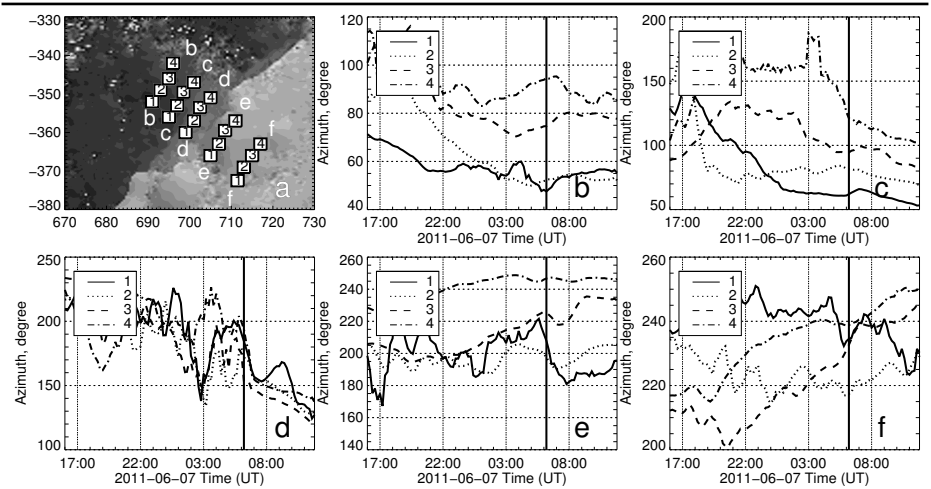}}
\caption{Panel (a) shows the magnetic field azimuth $\psi$ distribution on a part of the solar disk at $t=$ … UT (…) and the plotted squares of the spatial averaging for $\psi$. Panels (b - f) present the $\psi$ time variations within different squares.}
\end{figure}

\section{Discussion and conclusions}
     \label{S-Conclusion} 

The 2011 June 7 eruptive event involves a filament eruption, a flare, and a CME occurrence. Our analysis showed that this event started with FE, that was, apparently, the trigger for two other eruptions. Hence, it follows that, by revealing the relation between the  magnetic field variations in the eruptive event region and its components (FE, flare, and CME), we should, first of all, study, how the field variations are related to FE. The filament eruption at its all three stages identified by the time dependence of the filament velocity (Fig. 2) appeared not to be accompanied by value-significant emergences of long-living NMFs in the form of bipolar regions. The filament motion before the flare onset is accompanied by an eventual decrease in the unsigned magnetic flux in several sites of the photosphere in the neighboring of the filament channel that adjoin PIL from the different-polarity field (Fig. 3). Simultaneously, in some such sites, involving the different-polarity domains, the magnetic flux decrease within the different-polarity domains occurs at approximately equal rate. This gives rise to consider that, in such sites of the photosphere, one observes magnetic flux cancellation (MFC). This process, as already noted, may be an FE trigger. At the same time, we failed to reveal a physical association between this process and the eruptive filament motion peculiarities awhile.

The filament eruption appeared to be accompanied by another interesting phenomenon. During all the time while the filament was moving, Spot 8 in Fig. 3 moved sufficiently rapidly south-eastward. Before the flare onset, the spot mean velocity was $\approx$ 130 m/s. Before the flare, the spot size and ``blackening'' degree varied weakly, and after the flare onset, these characteristics of the spot began to vary substantially more intensely: the spot size decreased and its darkness faded. Also, the spot velocity decreased manyfold. We revealed that the south-eastern footpoint of the filament also moved south-eastward before the flare onset, but at a smaller velocity. One may assume that such a motion increases the flux rope shear, which, in turn, also raises chances to disbalance a filament \citep{Chen2011}. Probably, there is a relation between the spot and the filament footpoint motions, and one may consider the spot motion as an indicator for the filament shear increase.

The variation the photospheric magnetic field within the eruption region before the flare onset and during the flare featured different intensities and time scales. The flare is preceded by a set of rapid (with the time scale of, at least, 12 minutes = HMI temporal resolution at the vector measurements of the field) small-scale spatial variations in the magnetic field. These variations include the variations in all the magnetic induction components and in the neutral line. Simultaneously, the magnetic induction modulus decreased (against slight variations) during about 24 hours before the flare onset within the region of the group of the sunspots in the filament feet neighboring and in the vicinities of some single sunspots. One observes interesting field variations after the flare onset or during the eruptive filament velocity dramatic acceleration last stage that is close in time to the flare onset. In a part of the flare region in the PIL neighboring, the magnetic induction modulus $(B)$ increased by about $\approx 140 - 150 G$ for approximately 40 minutes after the flare onset. The comparison between the field variations and the variations in the transversal ($B_t$) and radial ($B_r$) magnetic induction components in this part showed that the $B$ increase occurs, mainly, due to the $B_t$ growth. The $B_r$ variation after the flare onset is insignificant, and the $B_r$ modulus persisted relatively small within 2-3 hours during the flare.

A consequence of such $B$, $B_t$, and $B_r$ variations is a dramatic (in about 40 minutes) increase in the domain of the magnetic induction modulus increased values $\alpha$ that is the field line inclination angle to the radial direction from the Sun center. The maximal increase in the angle $\alpha$ reached $25^\circ - 30^\circ$. As seen from Fig. 11, the strongest increases in the angles $\alpha$ occurred in the on-PIL squares, within which the angle $\alpha$ was averaged. But, exactly on PIL, $\alpha=90^\circ$, because this is the site, where $B_r=0$. Thereby, there is a following question: how can one elucidate the existence of angles $\alpha$ essentially smaller than $90^\circ$ in the on-PIL squares before the flare, and that dramatically increase to more than $80^\circ$ during the flare? One may elucidate these peculiarities of the angle $\alpha$ behavior in the PIL neighboring as follows. Close to PIL, the angle $\alpha$ dramatically decreases, when the square moves away from PIL. As a result, the mean value $\alpha$ in the squares used for averaging appears substantially smaller than $90^\circ$. During flares, angles $\alpha$, when moving away from PIL, increase, therefore, the $\alpha$ distribution in the PIL neighboring becomes smoother, which leads to a growth in the field line inclination angles, when averaging in the on-PIL squares.

We also studied the behavior of angle $\alpha$ as the moved away from PIL and from the domain, where the field line inclination angle increased during flares. Selected were the sites, in which one did not observe short-time emergence of the PIL fragments. The angle $\alpha$ appeared to decrease against the variations before the flare onset on the bulk of the sites under consideration. Such an $\alpha$ decrease may be elucidated through the filament rise and through the simultaneous extraction of the magnetic field lines encompassing the filament. Different was the character of the time variations in the $\alpha$ angles within the sites adjoining the south-eastern footpoint of the filament. On these sites, the $\alpha$ angle increased before the flare onset. The reason for that is, probably, the filament expansion along its axis, or, in other words, an increase in its longitudinal angular size resulting in a departure of the ambient field lines adjoining the filament feet.

We also studied the properties of the photospheric field in the flare ribbon region. In first approximation, irrespective of the diverging ribbon's position, the ribbons (in the projection onto the photosphere) coincide with the $B$ local maxima and $\alpha$ minima.
And, finally, note that we studied the variations in the magnetic field azimuth. We found that the time variations in this angle depend on the position of the site, where these variations are analyzed. In the PIL domain, the field azimuth after the flare onset dramatically decreased. This means that the angle between PIL and the projection of the magnetic induction vector onto the sky plane (we designate this angle $\beta$) also decreased. In the photoshere sites remote from the magnetic field PIL, one observes a slight growth in the azimuth after the flare onset, which means that the angle $\beta$, accounting for the azimuth values before the flare onset, increases.

%

%
 \begin{acks}
The authors thank the SDO/AIA, SDO/HMI, and GOES teams for a possibility to freely use the data from these instruments. The authors thank V.V. Grechnev, who provided them the software for the solar differential rotation compensation. The study was done with the support from Russian Foundation for Basic Research Grants No. 15-02-01077-а and No. 16-32-00315.
 \end{acks}

%
%
 \bibliographystyle{spr-mp-sola}
 \bibliography{biblio}

\end{article} 
\end{document}